\documentclass[a4paper,UKenglish]{lipics_md}
\usepackage{microtype}
\usepackage{graphicx}
\usepackage{tikz}
\usepackage{color}

\title{The Weighted Mean Curvature Derivative
       of a Space-Filling Diagram\footnote{
        This project has received funding from the European Research Council (ERC)
        under the European Union's Horizon 2020 research and innovation programme
        (grant agreement No 78818 Alpha).
	It is also partially supported by the DFG Collaborative Research Center TRR 109,
	`Discretization in Geometry and Dynamics',
	through grant no.\ I02979-N35 of the Austrian Science Fund (FWF).}}

\author[1]{Arsenyi Akopyan}
\author[1]{Herbert Edelsbrunner}
\affil[1]{IST Austria (Institute of Science and Technology Austria),
	Klosterneuburg, \\
        Austria, \texttt{edels@ist.ac.at}, \texttt{akopjan@gmail.com}}
\authorrunning{A. Akopyan and H. Edelsbrunner}

\keywords{Molecular dynamics, proteins, space-filling diagrams, intrinsic volume,
  alpha shapes, inclusion-exclusion, derivatives, discontinuities, computer implementation.}

\newcommand {\mm}[1] {\ifmmode{#1}\else{\mbox{\(#1\)}}\fi}

\newcommand {\scalprod}[2] {{\langle #1 , #2 \rangle}}
\newcommand{\denselist}{\itemsep 0pt\parsep=1pt\partopsep 0pt}

\newcommand{\ourproof}{\begin{proof}}
\newcommand{\eop}{\end{proof}}

\newcommand{\Bspace}       {\mm{{\mathbb B}}}
\newcommand{\Mspace}[1]    {\mm{{\mathbb M}_{\rm {#1}}}}
\newcommand{\Rspace}       {\mm{{\mathbb R}}}
\newcommand{\Sspace}       {\mm{{\mathbb S}}}
\newcommand{\Grass}[2]     {\mm{{\mathcal G}_{#1}^{#2}}}

\newcommand{\power}[1]     {\mm{\pi_{#1}}}
\newcommand{\weight}[1]    {\mm{{w_{#1}}}}
\newcommand{\Vdom}[1]      {\mm{{V}_{#1}}}
\newcommand{\Volume}       {\mm{\it Volume}}
\newcommand{\Area}         {\mm{\it Area}}
\newcommand{\Length}       {\mm{\it Length}}
\newcommand{\Angle}        {\mm{\it Angle}}
\newcommand{\Card}[1]      {\mm{\#}{({#1})}}

\newcommand{\volume}       {\mm{\it vol}}
\newcommand{\area}         {\mm{\it area}}
\newcommand{\length}       {\mm{\it length}}
\newcommand{\mean}         {\mm{\it mean}}
\newcommand{\gauss}        {\mm{\it gauss}}

\newcommand{\Euler}[1]     {\mm{\chi}{[{#1}]}}

\newcommand{\interior}[1]  {\mm{\rm int\,}{#1}}
\newcommand{\boundary}[1]  {\mm{\rm bd\,}{#1}}
\newcommand{\diff}         {\mm{\rm d}}
\newcommand{\intdiff}      {\mm{\rm \,d}}
\newcommand{\Diff}         {\mm{\rm D}}
\newcommand{\norm}[1]      {\mm{\|{#1}\|}}

\newcommand{\Edist}[2]     {\mm{\|{#1}-{#2}\|}}
\newcommand{\Fdist}[2]     {\mm{\|{#1}\!-\!{#2}\|}}

\newcommand{\Aalpha}[1]    {\mm{\alpha}^{#1}}
\newcommand{\aaa}          {\mm{\bf a}}
\newcommand{\bbb}          {\mm{\bf b}}
\newcommand{\ccc}          {\mm{\bf c}}
\newcommand{\ddd}          {\mm{\bf d}}

\newcommand{\mmm}          {\mm{\bf m}}
\newcommand{\ooomega}      {\mm{\boldsymbol\omega}}
\newcommand{\ppp}          {\mm{\bf p}}
\newcommand{\qqq}          {\mm{\bf q}}
\newcommand{\sss}          {\mm{\bf s}}
\newcommand{\ttt}          {\mm{\bf t}}
\newcommand{\tttStr}       {\mm{\bf t}^{\rm str}}
\newcommand{\tttRot}       {\mm{\bf t}^{\rm rot}}
\newcommand{\uuu}          {\mm{\bf u}}
\newcommand{\vvv}          {\mm{\bf v}}
\newcommand{\xxx}          {\mm{\bf x}}
\newcommand{\TTT}          {\mm{\bf T}}
\newcommand{\UUU}          {\mm{\bf U}}
\newcommand{\VVV}          {\mm{\bf V}\!}
\newcommand{\Welec}        {\mm{W_{\!{\rm elec}}}}
\newcommand{\Wnp}          {\mm{W_{\!{\rm np}}}}

\newcommand{\ee}           {\mm{\varepsilon}}
\newcommand{\Case}[2]      {\mm{{\bf {#1}}_{\bf {#2}}}}

\newcommand{\ourparagraph}[1] {\vspace{0.1in} \noindent \textbf{#1}}
\newcommand{\Skip}[1]      {}

\begin{document}
\maketitle

\begin{abstract}
  Representing an atom by a solid sphere in $3$-dimensional Euclidean space,
  we get the space-filling diagram of a molecule by taking the union.
  Molecular dynamics simulates its motion subject to bonds and other forces,
  including the solvation free energy.
  The morphometric approach \cite{HRC13,RHK06} writes the latter
  as a linear combination of weighted versions of the volume, area,
  mean curvature, and Gaussian curvature of the space-filling diagram.
  We give a formula for the derivative of the weighted mean curvature.
  Together with the derivatives of the weighted volume in \cite{EdKo03},
  the weighted area in \cite{BEKL04}, and the weighted Gaussian curvature \cite{AkEd19},
  this yields the derivative of the morphometric expression of the
  solvation free energy.
\end{abstract}

\section{Introduction}
\label{sec:1}

This paper makes a significant step toward turning the
morphometric approach to modeling the solvation free energy in molecular dynamics
into practice.
We recall that \emph{molecular dynamics} uses Newton's second law of motion
to simulate the dynamic behavior of molecules.
We focus on the case in which the motion is computed within an
\emph{implicit solvent model}, in which the effect of water is captured by
an effective solvation potential, $W = \Welec + \Wnp$.
The first term on the right-hand side accounts for the electrostatic polarization,
and the second term for the van der Waals interactions and the formation
of a void in the solvent.
Referring to a large body of work on $\Welec$ \cite{Sim03},
we focus on the \emph{non-polar} or \emph{hydrophobic effect} of water,
namely $\Wnp$.
In an effort to quantify this effect,
Lee and Richards introduced the \emph{solvent-accessible surface}
of a molecule \cite{LeRi71}.
It is the boundary of the space-filling diagram in which each atom is
represented by a ball whose radius is its van der Waals radius plus
$1.4$ Angstrom for the approximate radius of a water molecule.
Based on this concept, Eisenberg and McLachlan defined the
\emph{solvation free energy} as a weighted sum of the surface area,
$\Wnp = \sum_i \weight{i} A_i$,
in which $\weight{i}$ is the atomic solvation parameter and $A_i$
is the area of the sphere modeling the $i$-th atom that is
accessible to the solvent \cite{EiMc86}.
There is, however, disagreement about the precise formulation fueled in part
by evidence that for a small solute, $\Wnp$ is more closely related to
the solvent-excluded volume rather than the solvent-accessible area \cite{LCW99,SiBr94}.
To resolve this controversy, the \emph{morphometric approach} of Mecke and Roth
\cite{HRMD07,HRC13,KRM04,Mec96,RHK06} suggests that $\Wnp$ be a linear combination of
the intrinsic volumes, which in $\Rspace^3$ are
the volume, area, mean curvature, and Gaussian curvature.
This approach is inspired by the mathematical theory of intrinsic volumes,
which we now summarize.

To begin, we let $K \subseteq \Rspace^3$ be a compact convex body
and write $K_r = K \oplus r \Bspace^3$ for the \emph{parallel body} obtained
by thickening $K$ by $r$ all around.
As proved by Jakob Steiner, the volume of the parallel body is described by
a degree-$3$ polynomial,
\begin{align}
  \Volume(K_r)  &=  a_0 + a_1 r + a_2 r^2 + a_3 r^3 ,
  \label{eqn:Steiner}
\end{align}
in which $a_0$, $a_1$, $a_2$, and $a_3$ are the volume, surface area,
mean curvature, and one third of the Gaussian curvature of $K$ \cite{Ste40}.
Scaled and re-indexed versions of the coefficients are referred to as the
\emph{intrinsic volumes} of $K$.
According to Hugo Hadwiger's characterization theorem,
every measure of $K$ that is invariant under rigid motions, continuous, and additive
is a linear combination of the intrinsic volumes \cite{Had51}.
Since the solvent-accessible model is obtained by thickening the van der Waals model,
the analogy seems compelling, except that molecules are generally not convex.
To bridge this gap, we mention a result by Morgan Crofton,
which states that the $i$-th coefficient of the Steiner polynomial satisfies
\begin{align}
  a_i  &=  c_i \int_{H \in \Grass{i}{3}}
               \int_{x \perp H} \Euler{K \cap (H + x)} \intdiff x \intdiff H ,
  \label{eqn:Crofton}
\end{align}
in which $c_0 = 1$, $c_1 = \tfrac{2}{\pi}$, $c_2 = 1$, $c_3 = \tfrac{4 \pi}{3}$
\cite{Cro68}.
To explain \eqref{eqn:Crofton}, we note that $\Grass{i}{3}$ is the
\emph{Grassmannian} that consists of all $i$-planes passing through
the origin in $\Rspace^3$, $H$ is such an $i$-plane,
$x$ is a point on the orthogonal $(3-i)$-plane,
$H+x$ is the $i$-plane parallel to $H$ that passes through $x$,
and $\chi$ is the Euler characteristic.
Importantly, Crofton's integral formula \eqref{eqn:Crofton} is not restricted to convex bodies
and can thus be used to extend the definition of intrinsic volume to non-convex bodies.
Returning to the modeling of molecules, we embrace the morphometric approach.
In other words, we write
\begin{align}
  \Wnp  &=  \mu_0 \cdot a_0 + \mu_1 \cdot a_1 + \mu_2 \cdot a_2 + \mu_3 \cdot a_3 ,
  \label{eqn:morphometric}
\end{align}
in which the $a_i$ are the volume, area, mean curvature,
and one third of the Gaussian curvature of the solvent-accessible model,
and physical meanings of the $\mu_i$ can be found in \cite{HRC13,RHK06}.
To model physical properties, such as hydrophobicity, we consider
weighted versions of the four measures.
To embed this approach into molecular dynamics software,
we need the derivative of $\Wnp$
with respect to the motion or, equivalently, the derivatives of the
weighted volume, the weighted area, the weighted mean curvature,
and the weighted Gaussian curvature.
The first two of those derivatives have been studied in \cite{EdKo03} and \cite{BEKL04},
where we find formulas based on the Voronoi decomposition of the space-filling diagram
as well as characterizations of the configurations at which the derivative
is not continuous.
The main result of this paper is a similar analysis of the weighted mean curvature
derivative, and we refer to \cite{AkEd19} for the last piece of the puzzle,
which is the weighted Gaussian curvature derivative.

\ourparagraph{Outline.}
Section \ref{sec:2} reviews the background relevant to this paper.
Section \ref{sec:3} derives the constituents of the weighted mean
curvature of a space-filling diagram.
Section \ref{sec:4} states the derivative in terms of the gradient.
Section \ref{sec:5} analyzes the configurations at which the gradient
is not continuous.
Section \ref{sec:6} concludes this paper.

\section{Background}
\label{sec:2}

The background needed for the technical results reported in this paper include
two classic theorems in geometry attributed to Archimedes and to Heron of Alexandria,
the concept of a space-filling diagram to represent a molecule,
the corresponding alpha complex to do the book-keeping,
and the analysis of the weighted volume and weighted area derivatives,
which are heavily based on these concepts.

\ourparagraph{Two area formulas.}
Write $\Sspace^2$ for the unit sphere in $\Rspace^3$
and recall that its area is $4 \pi$.
Accordingly, the area of a sphere with radius $r$ is $4 \pi r^2$.
Spheres in $\Rspace^3$ enjoy a property discovered thousands of
years ago by Archimedes, which is unique to three dimensions.
To describe it, we use Cartesian coordinates and let
$f \colon \Sspace^2 \to \Rspace$ map every point to its third coordinate,
which we call its \emph{height}.
By Archimedes, the area of the cap consisting of all points at height at
most $-1 \leq a \leq 1$ depends linearly on $a$
and is therefore $4 \pi$ times $1+a$ divided by the diameter of
the unit sphere, which is $2$.
\begin{formula}[Archimedes' Theorem]
  \label{form:Archimedes}
  For every $-1 \leq a \leq 1$, the area of the cap $f^{-1} [-1, a] \subseteq \Sspace^2$
  is $2 \pi (1+a)$.
\end{formula}
A slice of the sphere can be written as the preimage of an interval,
$[a,b] \subseteq [-1,1]$.
The area of $f^{-1} [a,b]$ is the area of $f^{-1} [-1,b]$
minus the area of $f^{-1} [-1,a]$, which is $2 \pi (b-a)$.
We note that the area of the slice thus depends on the width, $b-a$,
but is independent of the location of the interval within $[-1,1]$.

A formula that gives the area of a triangle in terms of the lengths
of the three edges is attributed to Heron of Alexandria,
although it has been suggested that Archimedes knew about the formula
two centuries earlier.
\begin{formula}[Heron's Theorem]
  \label{form:Heron}
  The area of a triangle with edges of length $a, b, c$ is
  $A = \sqrt{ s (s-a) (s-b) (s-c)}$, in which $s = \tfrac{1}{2}(a+b+c)$.
\end{formula}
There are various proofs of this formula and a number of equivalent statements.
We will use $A = \tfrac{1}{4} \sqrt{ 2 (a^2b^2 + b^2c^2 + a^2c^2) - (a^4 + b^4 + c^4)}$.

\ourparagraph{Space-filling diagrams.}
Mathematically, such a diagram is merely the union of finitely many
closed balls in $\Rspace^3$.
Its significance derives from the use as a geometric model of
biomolecules \cite{LeRi71}, such as proteins, DNA, and the like.
There are competing models, such as the level sets of a sum
of isotropic Gaussian distributions,
but we focus on the geometrically more concrete space-filling diagrams.
This section introduces the concept along with most
of the notation needed for its discussion in the technical sections.

Throughout this paper, we assume a set of $n$ closed balls in $\Rspace^3$,
$X = \{ B_i \mid 0 \leq i < n \}$,
in which $x_i \in \Rspace^3$ is the center and $r_i \in \Rspace$
is the radius of $B_i$.
The \emph{space-filling diagram} is the union of these balls,
$\bigcup X = \bigcup_{i=0}^{n-1} B_i$.
We are interested in the volume, area, mean curvature, Gaussian curvature
of this diagram.
The volume should be clear.
To describe the other three measures, we note that the boundary
of $\bigcup X$ consists of patches of spheres that remain after
we remove open caps,
$S_i \setminus \bigcup_{j \neq i} \interior{B_j}$,
in which $S_i = \boundary{B_i}$.
These patches meet along portions of circles that remain after
we remove open arcs,
$S_{ij} \setminus \bigcup_{k \neq i,j} \interior{B_k}$,
in which $S_{ij} = S_i \cap S_j$.
These portions of circles consist of closed arcs that meet
at corners of the form
$S_{ijk} \setminus \bigcup_{\ell \neq i,j,k} \interior{B_\ell}$,
in which $S_{ijk} = S_i \cap S_j \cap S_k$.
For ease of reference, we summarize and extend the introduced notation
in Table \ref{tbl:Notation}, which is given in the appendix.
The area is simply the sum of areas of the sphere patches.
The mean curvature is more subtle because the boundary of the
space-filling diagram is not a smooth surface.
Instead of defining it as an integral of the point-wise mean curvature,
we use a discrete formula according to which the contribution of an
arc in the boundary of $\bigcup X$ is its length times
the dihedral angle between the outward normals of the two spheres that share the arc;
see \eqref{eqn:mean} for the formula in the weighted case.
Similarly, we write the Gaussian curvature as a sum of
contributions of sphere patches, circular arcs, and corners;
see \eqref{eqn:gauss} for the formula in the weighted case.
Alternatively, we can make use of the Gauss--Bonnet theorem
according to which the Gaussian curvature is $2 \pi$ times
the Euler characteristic of the surface,
but no such shortcut is available for the weighted case,
which we discuss next.

\ourparagraph{Voronoi decomposition and dual alpha complex.}
To generalize the measures to the weighted case,
we introduce the Voronoi decomposition of the space-filling diagram
and its dual, the alpha complex,
which serves as a book-keeping device.

Recall that $X$ is a set of $n$ closed balls $B_i$ with
centers $x_i$ and radii $r_i$.
The \emph{power distance} of a point $a \in \Rspace^3$
from $B_i$ is $\power{i} (a) = \Edist{a}{x_i}^2 - r_i^2$,
and we note that $B_i = \power{i}^{-1} (- \infty, 0]$.
The \emph{Voronoi domain} of $B_i$
is $\Vdom{i} = \{ a \in \Rspace^3 \mid \power{i} (a) \leq \power{j} (a)
    \mbox{\rm ~for~} 0 \leq j < n \}$,
and we write $\Vdom{ij}$, $\Vdom{ijk}$, and $\Vdom{ijk\ell}$
for the pair-wise, triple-wise, and quadruple-wise intersections.
The Voronoi domains decompose the space-filling diagram into
convex sets of the type $B_i \cap \Vdom{i}$.
We will need notation for the fraction of the ball represented
by this set,
and similarly for the fractions of common intersections,
which we collect in Table \ref{tbl:Notation}, which is given in the appendix.
Letting $\weight{i} \in \Rspace$ be the \emph{weight} of the $i$-th closed ball,
the formulas for the weighted measures of a space-filling diagram are given in
\eqref{eqn:volume}, \eqref{eqn:area}, \eqref{eqn:mean}, and \eqref{eqn:gauss} below.

The fractional measures listed in Table \ref{tbl:Notation}
are conveniently computed using the \emph{alpha complex} of $X$,
which is the nerve of the sets $B_i \cap \Vdom{i}$ \cite{EdMu94}.
Recall that the Delaunay mosaic is the nerve of the sets $\Vdom{i}$,
which implies that the alpha complex is a subcomplex of the
Delaunay mosaic,
and generically, both are simplicial complexes realized in $\Rspace^3$.
It will be convenient to denote a simplex by the set of indices of its vertices,
or by the sequence if we need an ordered version of the simplex.
By \emph{alpha shape} we mean the underlying space of the alpha complex.
To explain the connection between the simplices and the measures,
we note that a tetrahedron $ijk\ell$ in the Delaunay mosaic
belongs to the alpha complex iff $\nu_{ijk\ell} > 0$,
and the same rule applies to the triangles, edges, and vertices
of the mosaic.
Furthermore, a triangle $ijk$ in the alpha complex belongs to the
boundary of the alpha shape iff $\sigma_{ijk} > 0$,
and the same rule applies to the edges and vertices of the complex.
To compute these fractions, we use inclusion-exclusion over
subsets of simplices in the alpha complex.
For example,
\begin{align}
  \sigma_i  &=  1 - \tfrac{\sum_j \Area(B_j \cap S_i)}
                         {\Area(S_i)}
                  + \tfrac{\sum_{j,k} \Area(B_j \cap B_k \cap S_i)}
                         {\Area(S_i)}
                  - \tfrac{\sum_{j,k,\ell} \Area(B_j \cap B_k \cap B_\ell \cap S_i)}
                         {\Area(S_i)} ,
\end{align}
in which the sums are over all edges, triangles, tetrahedra
in the alpha complex that share $x_i$ as a vertex.
Proofs and additional formulas can be found in \cite{Ede95}.

\ourparagraph{Weighted intrinsic volume.}
Recall that $\bigcup X$ is the union of the balls $B_i$, for $0 \leq i < n$.
Its \emph{state}, $\xxx \in \Rspace^{3n}$, is the concatenation of the center vectors.
In other words, the $(3i+\ell)$-th coordinate of $\xxx$ is the $\ell$-th coordinate
of $x_i$.
We use the Voronoi tessellation to divide the space-filling diagram
into individual contributions and get the weighted intrinsic volume
by multiplication with the corresponding weight.
\begin{proposition}[Weighted Intrinsic Volumes]
  \label{prop:IntrinsicVolumes}
  The weighted volume, surface area, mean curvature, and Gaussian curvature
  of the space-filling diagram, $\bigcup X$, are
  \begin{align}
    \volume (\xxx) &=  \frac{4 \pi}{3} \sum_i \weight{i} \nu_i r_i^3,
      \label{eqn:volume} \\
    \area (\xxx)   &=  4 \pi \sum_i \weight{i} \sigma_i r_i^2 ,       
      \label{eqn:area}   \\
    \mean (\xxx)   &=  4 \pi \sum_i \weight{i} \sigma_i r_i
                       - \frac{\pi}{2} \sum_{i,j} (\weight{i}+\weight{j})
                                             \sigma_{ij} \phi_{ij} r_{ij} ,     
      \label{eqn:mean}   \\
    \gauss (\xxx)  &=  4 \pi \sum_i \weight{i} \sigma_i 
                       - \frac{\pi}{2} \sum_{i,j} (\weight{i}\!+\!\weight{j})
                                             \sigma_{ij} \lambda_{ij}
                     + \frac{1}{3} \sum_{i,j,k}
                          (\alpha_i \weight{i} \!+\! \alpha_j \weight{j}
                         \!+\! \alpha_k \weight{k}) \sigma_{ijk} \phi_{ijk} .
      \label{eqn:gauss}
  \end{align}
\end{proposition}
To get the above formula for the mean curvature, we smoothen the crevices of the
space-filling diagram by rolling a ball of radius $\ee > 0$ about the surface.
For sufficiently small $\ee$, the resulting surface is everywhere differentiable,
and we take the limit of its total mean curvature as $\ee$ goes to zero.
Along circular arcs, we partition the mean curvature in equal parts to the
two intersecting spheres.
The rationale for this rule is that the surface swept out by the centers of the
rolling balls of different radii partitions the exterior dihedral angle
at the arc in equal parts.
Using the same geometric intuition, we partition the contribution of a corner
of the space-filling diagram to the Gaussian curvature in proportions
$\alpha_i + \alpha_j + \alpha_k = 1$;
see \cite{AkEd19} for details.
All variables have been defined above,
except for $\phi_{ij}$ (the angle between the unit normals
of the spheres at a point of $S_{ij} = S_i \cap S_j$),
$\lambda_{ij}$ (the combined length of the two normals
after projection to the line passing through $x_i$ and $x_j$),
and $\phi_{ijk}$ (the solid angle spanned by the unit normals of
$S_i$, $S_j$, $S_k$ at a point of common intersection, $P \in S_{ijk}$).

\ourparagraph{Weighted volume and area derivatives.}
To state the derivative of the weighted volume,
we introduce the \emph{momentum}, $\ttt \in \Rspace^{3n}$,
which is the concatenation of the velocity vectors.
In other words, the vector $\ttt_i = [t_{3i+1}, t_{3i+2}, t_{3i+3}]^T$
is the velocity vector of the $i$-th ball.
The derivative of a function $F \colon \Rspace^{3n} \to \Rspace$
can be given in terms of the gradient:
$\Diff F_\xxx (\ttt) = \scalprod{\nabla_{\!\xxx} f}{\ttt}$.
Writing $\vvv = \nabla_{\!\xxx} \volume$ and
$\vvv_i = [v_{3i+1}, v_{3i+2}, v_{3i+3}]^T$,
\cite{EdKo03} gives the derivative
of the weighted volume:
\begin{proposition}[Gradient of Weighted Volume]
  \label{prop:VolumeDerivative}
  The derivative of the weighted volume of the space-filling diagram
  at state $\xxx$ with momentum $\ttt$ is
  $\Diff \volume_\xxx (\ttt) = \scalprod{\vvv}{\ttt}$ with
  \begin{align}
    \vvv_i  &=  \sum_j \left[ (v_{ij}+v_{ji}) \uuu_{ij}
                            + (\bar{v}_{ij}-\bar{v}_{ji}) \UUU_{ij} \right] , \\
    v_{ij}  &=  \tfrac{1}{2} \pi \weight{i} r_{ij}^2 \nu_{ij}
                \left( 1 + \frac{r_i^2 - r_j^2}{\Edist{x_i}{x_j}^2} \right) , \\
    \bar{v}_{ij}  &=  \tfrac{2}{3} \pi \weight{i} r_{ij}^2 \nu_{ij} \frac{1}{\Edist{x_i}{x_j}} ,
  \end{align}
  in which $\UUU_{ij}$ is the vector from $x_{ij}$ to the centroid
  of $B_{ij} \cap \Vdom{ij}$, and
  the sum is over all edges in the alpha complex that are incident to $x_i$.
\end{proposition}
Writing $\aaa = \nabla_{\!\xxx} \area$ and
$\aaa_i = [a_{3i+1}, a_{3i+2}, a_{3i+3}]^T$, \cite{BEKL04} gives the derivative
of the weighted area:
\begin{proposition}[Gradient of Weighted Area]
  \label{prop:AreaDerivative}
  The derivative of the weighted area of the space-filling diagram
  at state $\xxx$ with momentum $\ttt$ is
  $\Diff \area_\xxx (\ttt) = \scalprod{\aaa}{\ttt}$ with
  \begin{align}
    \aaa_i  &=  \sum_j (a_{ij}+a_{ji}) \uuu_{ij}
            + \sum_{j,k} (a_{ijk}-a_{jik}) \uuu_{ijk} , \\
    a_{ij}  &=  \pi \weight{i} \sigma_{ij} r_i
                \left(1 - \frac{r_i^2 - r_j^2}{\Edist{x_i}{x_j}^2} \right)        , \\
    a_{ijk}  &=  2 \weight{i} r_{ijk} \nu_{ijk} r_i \frac{1}{\Edist{x_i}{x_j}}    ,
  \end{align}
  in which the first sum is over the boundary edges in the alpha complex
  incident to $x_i$,
  and the second sum is over the triangles incident to these edges.
\end{proposition}

\section{Derivatives}
\label{sec:3}

We recall \eqref{eqn:mean} and decompose the weighted mean curvature function
into a first term, which accounts for the curvature within the sphere patches,
and a second term, which accounts for the curvature concentrated
along the circular arcs:
\begin{align}
  \mean (\xxx)  &=  4 \pi \sum_i \weight{i} r_i \sigma_i 
                  - \tfrac{\pi}{2} \sum_{i,j}  (\weight{i}+\weight{j})
                        r_{ij} \phi_{ij} \sigma_{ij} .
  \label{eqn:mean2}
\end{align}
While not reflected in the notation,
$\sigma_i$, $r_{ij}$, $\phi_{ij}$, and $\sigma_{ij}$ depend on the state.
It is convenient to write $\sigma_i (\tau)$ for $\sigma_i$
at state $\xxx + \tau \ttt$, and similarly for the other functions.
Furthermore, we write $\sigma_i'$ for the derivative of $\sigma_i$
at $\tau = 0$,
and similarly we write $r_{ij}'$, $\phi_{ij}'$, $\sigma_{ij}'$.
Derivatives with respect to parameters other than $\tau$
are explicitly stated as such.

\subsection{Derivative of $\sigma_i$}
\label{sec:31}

\begin{figure}[hbt]
  \centering \resizebox{!}{1.3in}{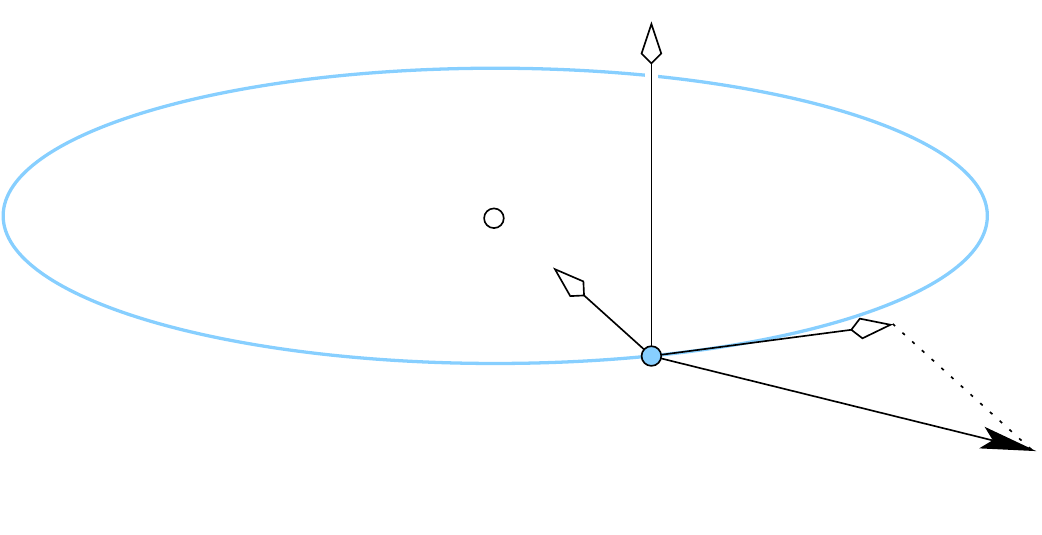}
  \caption{The angular momentum of the motion of $x_j$ relative to $x_i$.
    Projecting $\ttt_j - \ttt_i$ onto the tangent vector of the circular
    orbit, we get a system of pairwise orthogonal vectors,
    $\tttRot_j$, $\uuu_{ij}$, and $\ooomega_j$.}
  \label{fig:Momentum}
\end{figure}

To derive $\sigma_i$, we follow \cite{BEKL04} and decompose the motion into
a direction preserving component and a distance preserving component.
The direction preserving motion \emph{stretches} the distance between two
centers, while the distance preserving motion \emph{rotates} one center
about the other.
We therefore write $\ttt_j - \ttt_i = \tttStr_{j} + \tttRot_{j}$,
in which $\tttStr_{j} = \scalprod{\ttt_j-\ttt_i}{\uuu_{ij}} \uuu_{ij}$.
Observe that $\tttRot_{j} = (\ttt_j-\ttt_i) - \tttStr_{j}$
is the velocity vector of the rotation with \emph{angular momentum}
\begin{align}
  \ooomega_{j}  &=  \frac{(\ttt_j-\ttt_i) \times \uuu_{ij}}
                         {\Edist{x_i}{x_j}}           
                 =  \frac{\tttRot_{j} \times \uuu_{ij}}
                         {\Edist{x_i}{x_j}} ,
  \label{eqn:omega}
\end{align}
in which $\times$ is the \emph{cross} or \emph{outer product}
that maps vectors $a$ and $b$ to the vector $c = a \times b$
of length $\norm{a} \norm{b} \sin \angle a b$
normal to both in such a way that the sequence $a, b, c$ forms
a right-handed system in space;
see Figure \ref{fig:Momentum}.
The analysis of the direction preserving motion is similar
to the analysis of the area derivative in \cite{BEKL04},
and we repeat the main steps for completeness.
Since little in terms of a proof of the distance preserving motion
is offered in \cite{BEKL04}, we give a complete argument,
which can be translated back almost verbatim to a proof in the area case.

\ourparagraph{Direction preserving motions.}
Consider two balls, $B_i$ and $B_j$, and the change of surface area
under a motion of $x_i$ that preserves the direction, $\uuu_{ij}$.
Let $\Edist{x_i}{x_j} = \xi_i + \xi_j$,
in which the terms on the right-hand side of the equation are the
signed distances from the centers to the plane bisector.
As noted in \cite{BEKL04}, we have
\begin{align}
  \xi_i  &=  \frac{1}{2} \left( \Edist{x_i}{x_j} 
           + \frac{r_i^2 - r_j^2}{\Edist{x_i}{x_j}} \right) , 
  \label{eqn:xii}
\end{align}
and similarly for $\xi_j$.
Plugging \eqref{eqn:xii} into Formula \ref{form:Archimedes},
we get
\begin{align}
  \sigma_i  &=  \frac{r_i + \xi_i}{2 r_i}
             =  \frac{1}{4 r_i} \left( 2 r_i + \Edist{x_i}{x_j}
                  + \frac{r_i^2 - r_j^2}{\Edist{x_i}{x_j}} \right) .
\end{align}
Differentiating with respect to the distance between the two centers,
we get the rate of area change,
which happens right next to the circle of intersection.
In the general case, the cap $B_j \cap S_i$ may overlap with other caps so
that only a fraction of the bounding circle belongs to the boundary of
the space-filling diagram.
Accordingly, the derivative is the same fraction of the derivative
without any such overlap:
\begin{align}
  \frac{\diff \sigma_i}{\diff \Edist{x_i}{x_j}} 
            &=  \frac{\sigma_{ij}}{4 r_i}
                     \left( 1 - \frac{r_i^2 - r_j^2}{\Edist{x_i}{x_j}^2} \right).
  \label{eqn:diff1}
\end{align}
Multiplying the right-hand side of \eqref{eqn:diff1}
with $\scalprod{\ttt_j-\ttt_i}{\uuu_{ji}}$,
we get the derivative with respect to the direction preserving motion of $x_j$,
and taking the sum over all $j \neq i$, we get the derivative of $\sigma_i$
with respect to the direction preserving motions of all $x_j$;
compare with the first sum on the right-hand side of \eqref{eqn:dSigmai}
in Lemma \ref{lem:dSigmai}.

\begin{figure}[hbt]
  \centering \resizebox{!}{2.8in}{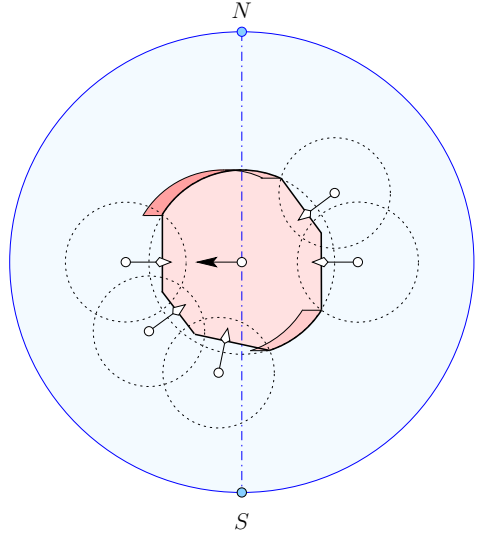}
  \caption{Blue sphere, $S_i$, with cap, $B_j \cap S_i$, in the middle
    moving due west.
    We shade the relevant part,
    which is the cap clipped to within the Voronoi domain of $x_j$.
    The motion divides its boundary into the front part (on the left)
    and the back part (on the right).
    Area loss and gain happen along the circular arcs bounding the clipped cap.
    The dotted caps surrounding $B_j \cap S_i$ are of the form $B_k \cap S_i$,
    and their vectors $\uuu_{ijk}$ are all orthogonal to
    the edge connecting $x_i$ with $x_j$.}
  \label{fig:Caps}
\end{figure}

\ourparagraph{Distance preserving motions.}
Consider first the case of a single cap, $B_j \cap S_i$.
The rotation defined by $\ooomega_{j}$
keeps the cap as well as $\sigma_i$ at constant size.
This is because $S_i$ loses area along the front of the circle that bounds the cap,
it gains area along the back of this circle, and the loss equals the gain.
In the more general case depicted in Figure \ref{fig:Caps},
it is possible that the loss and the gain no longer cancel each other.

To calculate the net loss or gain,
we let $N$ and $S$ be antipodal points on $S_i$, which we refer to as
\emph{north-} and \emph{south-pole}, such that the cap center
lies on the equator --- the great-circle halfway between $N$ and $S$ ---
and the motion follows the equator.
We recall that $\tttRot_{j}$ is the velocity vector of $x_j$,
and $r_i / \Edist{x_i}{x_j}$ times $\tttRot_{j}$
is the velocity vector of the cap center.
In Figure \ref{fig:Caps}, we draw $N\!S$ vertical,
we place $x_j$ and the cap center on top of $x_i$,
and we let the velocity vector go horizontally from right to left.
By Formula \ref{form:Archimedes}, the area loss along an arc of the
front does not depend on its position but only on the length of its
projection onto $N\!S$.
We are interested in arcs of $S_{ij}$ that lie on the boundary
of the space-filling diagram, and we compute the net loss or gain indirectly,
by measuring the projections of the edges in the Voronoi tessellation
defined by $B_i$, $B_j$, and a third ball, $B_k$.
Recall that $B_{ijk}$ is the line segment that connects the two points
in which $S_i, S_j, S_k$ meet,
and that $\nu_{ijk}$ is the fraction of this line segment that belongs
to the Voronoi decomposition of the space-filling diagram.
If this edge belongs to the front of the clipped cap, then its contribution
is positive because it subtracts from the loss along the front.
Symmetrically, if the edge belongs to the back of the clipped cap,
then its contribution is negative.
To quantify, we recall the definition of $\uuu_{ijk}$
given in Table \ref{tbl:Notation}.
The signed fraction of the projection of the relevant portion
of $B_{ijk}$ onto $N\!S$ is its length,
$2 r_{ijk} \nu_{ijk}$, times the cosine of the projection angle,
$\scalprod{\tttRot_{j}}{\uuu_{ijk}} / \norm{\tttRot_{j}}$,
divided by the diameter of the sphere, $2 r_i$.
By Formula \ref{form:Archimedes},
this is also the area fraction of the corresponding strip around the sphere.
Dividing by the length of the circular orbit,
$2 \pi \Edist{x_i}{x_j}$, we get the derivative:
\begin{align}
  \frac{\diff \sigma_i}{\diff \angle x_i x_j}
    &=  \sum_k \frac{r_{ijk} \nu_{ijk}}{2 \pi r_i \Edist{x_i}{x_j}}
        \scalprod{\tttRot_{j}}{\uuu_{ijk}} / \norm{\tttRot_{j}} ,
  \label{eqn:diff2}
\end{align}
in which we write $\angle x_ix_j$ for the angle that parametrizes
the rotation of $x_j$.
We get the contribution to $\sigma_i'$ by multiplying with $\norm{\tttRot_{j}}$
and observing that
$\scalprod{\tttRot_{j}}{\uuu_{ijk}} = \scalprod{\ttt_j-\ttt_i}{\uuu_{ijk}}$
for all $k$;
compare with the second sum on the right-hand side of \eqref{eqn:dSigmai}
in Lemma \ref{lem:dSigmai}.

\ourparagraph{Summary.}
A clipped cap whose entire boundary consists of straight line segments
does not touch the boundary of the space-filling diagram
and therefore has no contribution to the derivative.
These caps correspond to interior edges of the alpha complex.
Omitting them from the sum, we get the derivative of $\sigma_i$.
\begin{lemma}[Derivative of $\sigma_i$]
  \label{lem:dSigmai}
  The derivative of the fraction of $S_i$ on the boundary 
  of the space-filling diagram at state $\xxx$ with momentum $\ttt$ is
  \begin{align}
    \sigma_i'  &=  \sum_j \frac{\diff \sigma_i}{\diff \Edist{x_i}{x_j}}
                     \scalprod{\uuu_{ij}}{\ttt_i - \ttt_j}
                 + \sum_{j,k} \frac{r_{ijk} \nu_{ijk}}{2 \pi r_i \Edist{x_i}{x_j}}
                     \scalprod{\uuu_{ijk}}{\ttt_i - \ttt_j} , 
    \label{eqn:dSigmai} 
  \end{align}
  in which the first sum is over the boundary edges of the alpha complex incident
  to $x_i$, with coefficient given in \eqref{eqn:diff1},
  and the second sum is over the triangles incident to these edges.
\end{lemma}

\subsection{Derivatives of $r_{ij}$ and of $\phi_{ij}$}
\label{sec:32}

\begin{figure}[hbt]
  \centering \resizebox{!}{2.0in}{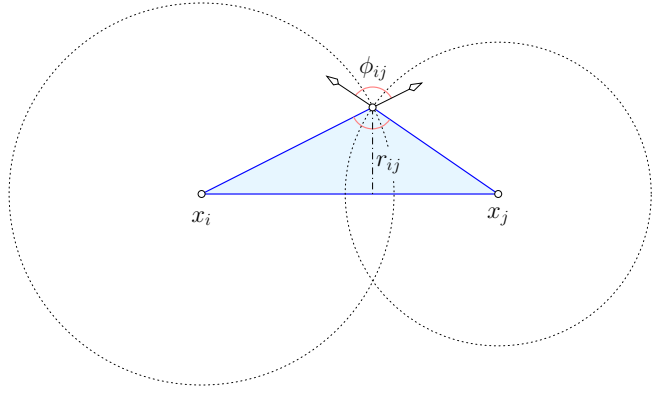}
  \caption{The radius of the circle, $r_{ij}$, in which the two spheres intersect
    is the height of the triangle whose base connects the centers of the spheres.
    Extending the two edges to the outward normals of the two spheres,
    we get the angle, $\phi_{ij}$, that is relevant in our computations.}
  \label{fig:Height}
\end{figure}

Recall that $r_{ij}$ is the radius of the circle $S_{ij} = S_i \cap S_j$,
assuming the two spheres have a non-empty intersection.
As illustrated in Figure \ref{fig:Height},
it is also the height of the triangle with base of length $\Edist{x_i}{x_j}$
and edges of length $r_i$ and $r_j$.
The area of the triangle is $A = \tfrac{1}{2} r_{ij} \Edist{x_i}{x_j}$.
Alternatively, we can compute the area using the version of the Heron formula
given right after Formula \ref{form:Heron}:
\begin{align}
  A  &=  \frac{1}{4}
         \sqrt{ 2 \left( r_i^2 r_j^2 + r_i^2 \Edist{x_i}{x_j}^2
                       + r_j^2 \Edist{x_i}{x_j}^2 \right)
                - \left( r_i^4 + r_j^4 + \Edist{x_i}{x_j}^4 \right) } .
  \label{eqn:Area}
\end{align}
The derivative of the area with respect to the distance between the two
centers is
\begin{align}
  \frac{\diff A}{\diff \Edist{x_i}{x_j}}
     &=  \frac{1}{32 A}
         \left( 4 \Edist{x_i}{x_j} (r_i^2+r_j^2) - 4 \Edist{x_i}{x_j}^3 \right) \\
     &=  \frac{1}{8 A} \Edist{x_i}{x_j} \left( r_i^2 + r_j^2 - \Edist{x_i}{x_j}^2 \right) .
\end{align}
From this, we get the derivative of the height of the triangle:
\begin{align}
  \!\!\frac{\diff r_{ij}}{\diff \Edist{x_i}{x_j}}
           &=  \frac{\diff \frac{2A}{\Edist{x_i}{x_j}}}
                    {\diff \Edist{x_i}{x_j}}
            =  \frac{2 \frac{\diff A}{\diff \Edist{x_i}{x_j}} \Edist{x_i}{x_j} - 2 A}
                    {\Edist{x_i}{x_j}^2}                                            \\
           &=  \frac{2 \Fdist{x_i}{x_j}^2
                     \left( r_i^2 +r_j^2 - \Fdist{x_i}{x_j}^2 \right) - 16 A^2}
                    {8 A \Fdist{x_i}{x_j}^2}                                
            =  \frac{\left( r_i^2 - r_j^2 \right)^2 - \Fdist{x_i}{x_j}^4}
                    {8 A \Fdist{x_i}{x_j}^2}                                         \\
           &=  \frac{\left( r_i^2 - r_j^2 \right)^2 - \Edist{x_i}{x_j}^4}
                    {2 \Fdist{x_i}{x_j}^2 
                       \sqrt{ 2 \left( r_i^2 r_j^2 + (r_i^2+r_j^2) \Fdist{x_i}{x_j}^2 \right) 
                       \!-\! \left( r_i^4 + r_j^4 + \Fdist{x_i}{x_j}^4 \right)}} .
  \label{eqn:rij}
\end{align}
Switching our attention, we observe that the angle between the outward normals,
$\phi_{ij}$, is also the angle opposite the base inside the triangle in
Figure \ref{fig:Height}.
Recall that the Law of Cosines generalizes Pythagoras' Theorem beyond
right-angled triangles: $c^2 = a^2 + b^2 - 2ab \cos \gamma$,
in which $a, b, c$ are the lengths of the sides and $\gamma$ is the
angle opposite to the side of length $c$.
In our application, we have $a = r_i$, $b = r_j$, $c = \Edist{x_i}{x_j}$,
and $\gamma = \phi_{ij}$.
Hence,
\begin{align}
  \phi_{ij}  &=  \arccos \frac{r_i^2 + r_j^2 - \Edist{x_i}{x_j}^2}
                              {2 r_i r_j} .
  \label{eqn:angle}
\end{align}
Recall that the derivative of $\arccos x$ is $-1 / \sqrt{1 - x^2}$.
Together with the chain rule for differentiation,
this gives the derivative of the angle with respect to the length of the base:
\begin{align}
  \frac{\diff \phi_{ij}}{\diff \Edist{x_i}{x_j}}
              &=  \frac{\frac{\Edist{x_i}{x_j}}{r_i r_j}}
                       {\sqrt{ 1 - \left( \frac{r_i^2 + r_j^2 - \Edist{x_i}{x_j}^2}
                                               {2 r_i r_j} \right)^2 }}       \\
              &=  \frac{2 \Edist{x_i}{x_j}}
                           {\sqrt{ 2 (r_i^2+r_j^2) \Edist{x_i}{x_j}^2
                                   - (r_i^2-r_j^2)^2 - \Edist{x_i}{x_j}^4 }} .
  \label{eqn:phiij}
\end{align}
We summarize the results of this subsection for later reference.
\begin{lemma}[Derivatives of $r_{ij}$ and of $\phi_{ij}$]
  \label{lem:dRijPhiij}
    The derivatives of the radius of the intersection circle of two spheres
    and of the angle between the outward normals at the intersection
    at state $\xxx$ with momentum $\ttt$ are
    \begin{align}
      r_{ij}'     &=  \frac{\diff r_{ij}}{\diff \Edist{x_i}{x_j}}
                      \scalprod{\uuu_{ij}}{\ttt_i-\ttt_j}
      \mbox{\rm ~~and~~}
      \phi_{ij}'   =  \frac{\diff \phi_{ij}}{\diff \Edist{x_i}{x_j}}
                      \scalprod{\uuu_{ij}}{\ttt_i-\ttt_j} ,
      \label{eqn:dRijPhiij}
    \end{align}
    with the coefficients in the two equations given in \eqref{eqn:rij} and \eqref{eqn:phiij}.
\end{lemma}

\subsection{Derivative of $\sigma_{ij}$}
\label{sec:33}

Recall that $\sigma_{ij}$ is the fraction of the circle $S_{ij} = S_i \cap S_j$
that belongs to the space-filling diagram.
We compute its derivative under the motion $\ttt$ in several steps,
the first of which modifies the motion.
Without altering the derivative, we do this such that the center and the plane
of the circle are fixed.

\ourparagraph{Modifying the motion.}
We begin by fixing $x_i$ in space, which we do by changing the velocity vector
of $x_j$ to $\ttt_j - \ttt_i$ for every $j$, as before.
Next, we fix the normal direction of the plane that contains $S_{ij}$
by removing the angular momentum.
Specifically, we change the velocity vector of $x_j$ to
$\VVV_{ij} = \ttt_j-\ttt_i - \ooomega_{j} \times  (x_j-x_i)$,
in which $\ooomega_j$ is given in \eqref{eqn:omega};
see Figure \ref{fig:Momentum}.
After this modification, the point $x_j$ moves with speed
$v$ along $\uuu_{ji}$, in which $v$ is the length of the new velocity vector.
Accordingly, we change the velocity vector of $x_k$ to
$\VVV_{ijk} = \ttt_k-\ttt_i - \ooomega_{j} \times (x_k-x_i)$ for every $k$,
and note that $\VVV_{iji} = 0$ and $\VVV_{ijj} = \VVV_{ij}$.
We finally fix the plane of the circle.
To this end, we recall that $\xi_i$ and $\xi_j$ given in \eqref{eqn:xii}
are the signed distances of $x_i$ and $x_j$ from the plane of $S_{ij}$.
We have $\xi_i + \xi_j = \Edist{x_i}{x_j}$ and use them to write
the center of the circle as an affine combination of the centers
of the spheres:
\begin{align}
  x_{ij}  &=  \frac{\xi_j}{\Edist{x_i}{x_j}} x_i
            + \frac{\xi_i}{\Edist{x_i}{x_j}} x_j                   \\
          &=  \frac{ \left( \Edist{x_i}{x_j}^2 - r_i^2 + r_j^2 \right) x_i
                   + \left( \Edist{x_i}{x_j}^2 + r_i^2 - r_j^2 \right) x_j }
                   { 2 \Edist{x_i}{x_j}^2 } .
  \label{eqn:xij}
\end{align}
Suppose now that $x_i$ is fixed and $x_j$ moves with speed $v$ in the
direction of $\uuu_{ji}$.
To compute the speed of $x_{ij}$ moving in the same direction,
we set $x_i = 0$, write $x_j = \Edist{x_i}{x_j}$,
and simplify \eqref{eqn:xij} to
$x_{ij} = \tfrac{1}{2} (\Edist{x_i}{x_j} + (r_i^2-r_j^2)/\Edist{x_i}{x_j})$.
Its derivative with respect to the distance is
\begin{align}
  \frac{\diff x_{ij}}{\diff \Edist{x_i}{x_j}}
    &=  \frac{1}{2} \left( 1 - \frac{r_i^2-r_j^2}{\Edist{x_i}{x_j}^2} \right) ,
  \label{eqn:D}
\end{align}
we write $D$ for the derivative in this special, $1$-dimensional scenario,
and note that $x_{ij}$ moves with speed $D v$.
Subtracting the corresponding multiple of $\uuu_{ji}$ from the velocity vector
of every point $x_k$, we get the final collection of vectors.
\begin{lemma}[Change of Motion]
  \label{lem:ReMotion}
  Replacing the velocity vector $\ttt_k$ by
  $\TTT_{ijk} = \VVV_{ijk} - D \VVV_{ij}$,
  for every $k$,
  fixes the center and the plane of the circle $S_{ij} = S_i \cap S_j$
  while preserving the derivative of $\sigma_{ij}$.
  The coefficient of $\VVV_{ij}$ is given in \eqref{eqn:D}.
\end{lemma}

\begin{figure}[hbt]
  \centering \resizebox{!}{2.0in}{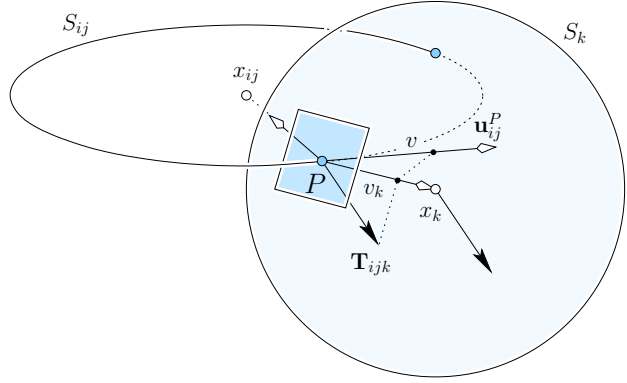}
  \caption{The sphere intersects the circle in two points,
    and we draw the tangent plane that touches the sphere in one of these points.
    The orthogonal projections of $\TTT_{ijk}$ and of $v \uuu_{ij}^P$
    onto $x_k - P$ give the speed $v_k$.}
  \label{fig:Arc}
\end{figure}

\ourparagraph{Movement of $x_k$.}
The circle $S_{ij}$ alternates between the boundary and the interior
of the space-filling diagram,
and we are interested in the arcs that belong to the boundary.
The endpoints of these arcs belong to $0$-spheres of the form
$S_{ijk} = S_i \cap S_j \cap S_k$.
Recall that $\sigma_{ijk}$ is $0$, $\tfrac{1}{2}$, or $1$.
If $\sigma_{ijk} = 1$, then both points of $S_{ijk}$ lie on the
boundary of the space-filling diagram and therefore serve as
endpoints of the relevant arcs of $S_{ij}$,
if $\sigma_{ijk} = \tfrac{1}{2}$,
then only one of the two points serves in this capacity,
and if $\sigma_{ijk} = 0$, then neither point serves in this capacity.
Consider a sphere, $S_k$, such that $\sigma_{ijk} = \tfrac{1}{2}$ or $1$,
and let $P \in S_{ijk}$ be an endpoint of an arc; see Figure \ref{fig:Arc}.
We are interested in the speed in which the point $P$ moves along $S_{ij}$.
To compute this speed, we let $\uuu_{ij}^P$ be the unit vector that is
tangent to $S_{ij}$ at the point $P$ such that $x_k$ and $P + \uuu_{ij}^P$
lie on the same side of the tangent plane, that is:
\begin{align}
  \uuu_{ij}^P  &=  \pm \frac{(x_{ij} - P) \times \uuu_{ij}}
                         {\Edist{x_{ij}}{P}} ,
\end{align}
and the correct sign is the one for which $\scalprod{x_k-P}{\uuu_{ij}^P} > 0$;
see Figure \ref{fig:Arc}.
When $x_k$ moves along $\TTT_{ijk}$ with speed $\norm{\TTT_{ijk}}$,
then the tangent plane moves along $x_k - P$ with speed
$v_k = \scalprod{x_k-P}{\TTT_{ijk}} / \Edist{x_k}{P}$.
At the same time, the tangent plane moves along $\uuu_{ij}^P$
with a speed $v$ such that substituting $v \uuu_{ij}^P$ for $\TTT_{ijk}$
in this equation gives the same speed, namely $v_k$.
In other words, $v \scalprod{x_k-P}{\uuu_{ij}^P} = \scalprod{x_k-P}{\TTT_{ijk}}$
and therefore $v = \scalprod{x_k-P}{\TTT_{ijk}} / \scalprod{x_k-P}{\uuu_{ij}^P}$.
This implies
\begin{align}
  \frac{\diff \sigma_{ij}}{\diff \angle x_ix_jx_k}
    &=  \frac{1}{\scalprod{x_k-P}{\uuu_{ij}^P}} 
        \scalprod{\TTT_{ijk}}{x_k-P} ,
\end{align}
in which $\angle x_ix_jx_k$ is the angle parametrizing the circular
motion of $x_k$ about the line passing through $x_i$ and $x_j$,
with $k$ is such that $P \in S_{ijk}$;
see the first sum on the right-hand side of \eqref{eqn:dRijPhiij}
in Lemma \ref{lem:dRijPhiij}.

\ourparagraph{Change of $r_{ij}$.}
Recall that Lemma \ref{lem:dRijPhiij} gives the rate of change of the radius.
We ask how it affects the derivative of $\sigma_{ij}$.
To begin, we observe that $h_k = \scalprod{x_{ij}-P}{x_k-P} / \Edist{x_k}{P}$
is the signed distance of $x_{ij}$ from the tangent plane.
It is positive if $x_{ij}$ and $x_k$ lie on the same side of the plane,
and negative if they lie on opposite sides.
Letting $\Aalpha{P}$ be the angle between $x_{ij}-P$ and the line in which
the tangent plane intersects the plane of $S_{ij}$,
we get $h_k = r_{ij} \sin \Aalpha{P}$.
Hence, $\Aalpha{P} = \arcsin (h_k / r_{ij})$,
and we are interested in the derivative with respect to the radius.
Remembering that the derivative of $\arcsin x$ is $1 / \sqrt{1 - x^2}$, we get
\begin{align}
  \frac{\diff \Aalpha{P}}{\diff r_{ij}}
    &=  \frac{1}{\sqrt{ 1 - h_k^2 / r_{ij}^2 }}
        \frac{-h_k}{r_{ij}^2}                                       
     =  \frac{- \scalprod{x_{ij}-P}{x_k-P}}
             {r_{ij} \sqrt{r_{ij}^2 \Edist{x_k}{P}^2 - \scalprod{x_{ij}-P}{x_k-P}^2} } .
  \label{eqn:alpha}
\end{align}
Using the chain rule, we get the derivative with respect to $\Edist{x_i}{x_j}$
by multiplying \eqref{eqn:alpha} with \eqref{eqn:rij} and then taking the sum
over all points $P$;
see the second sum on the right-hand side of \eqref{eqn:dRijPhiij} in Lemma \ref{lem:dRijPhiij}.

\ourparagraph{Summary.}
Adding the contribution of the changing radius
to those of the movements of the $x_k$ and normalizing by the
length of the circle, we get the desired derivative.
\begin{lemma}[Derivative of $\sigma_{ij}$]
  \label{lem:dSigmaij}
  The derivative of the fraction of $S_{ij}$ on the boundary
  of the space-filling diagram at state $\xxx$ with momentum $\ttt$ is
  \begin{align}
    \sigma_{ij}'  &=  \frac{1}{2 \pi r_{ij}} \left[
                      \sum_k \frac{\scalprod{\TTT_{ijk}}{x_k-P}}
                                  {\scalprod{x_k-P}{\uuu_{ij}^P}}
                    - \sum_k \frac{\diff \Aalpha{P}}{\diff r_{ij}}
                             \frac{\diff r_{ij}}{\diff \Edist{x_i}{x_j}}
                             \scalprod{\uuu_{ij}}{\ttt_j-\ttt_i} \right] ,
    \label{eqn:dsigmaij}
  \end{align}
  in which the coefficients in the second sum are given in \eqref{eqn:alpha}
  and \eqref{eqn:rij}.
  Both sums are over all oriented boundary triangles of the
  alpha shape that share the edge from $x_i$ to $x_j$,
  and $P \in S_{ijk}$ is the corresponding corner of the space-filling diagram.
\end{lemma}

\section{Gradients}
\label{sec:4}

We write the derivative of the weighted mean curvature function,
$\mean \colon \Rspace^{3n} \to \Rspace$,
in terms of the gradient of $\mean$ at $\xxx \in \Rspace^{3n}$,
denoted $\mmm = \nabla_{\!\xxx} \mean$.
Recalling \eqref{eqn:mean2}, this derivative is
$\mean' = p' + q' + s'$, with
\begin{align}
  p'  &=  4 \pi \sum_i \weight{i} r_i \sigma_i' ,
    \label{eqn:pprime} \\
  q'  &=  - \frac{\pi}{2} \sum_{i,j} (\weight{i}+\weight{j})
          \left( r_{ij}' \phi_{ij} \sigma_{ij} + r_{ij} \phi_{ij}' \sigma_{ij} \right),
    \label{eqn:qprime} \\
  s'  &=  - \frac{\pi}{2} \sum_{i,j} (\weight{i}+\weight{j}) r_{ij} \phi_{ij} \sigma_{ij}'.
    \label{eqn:sprime}
\end{align}
Writing $\mmm = [m_1, m_2, \ldots, m_{3n}]^T$,
we recall that $\mmm_i = [m_{3i+1}, m_{3i+2}, m_{3i+3}]^T$
is the $3$-dimensional gradient that applies to $x_i$.
Using boldface letters for the gradients of $p, q, s$,
and similar conventions for the $3$-dimensional sub-vectors,
we have $\mmm = \ppp + \qqq + \sss$ and
$\mmm_i = \ppp_i + \qqq_i + \sss_i$ for $0 \leq i < n$.
We get the gradients by redistributing the derivatives
stated in Lemmas \ref{lem:dSigmai} to \ref{lem:dSigmaij}.

\ourparagraph{First term.}
To begin, we use Lemma \ref{lem:dSigmai} to rewrite \eqref{eqn:pprime} as
\begin{align}
  p'  &=  \sum_{i,j}   p_{ij}  \scalprod{\uuu_{ij}}{\ttt_i-\ttt_j}
        + \sum_{i,j,k} p_{ijk} \scalprod{\uuu_{ijk}}{\ttt_i-\ttt_j} , 
    \label{eqn:First} \\
  p_{ij}  &=  4 \pi \weight{i} r_i \frac{\diff \sigma_i}{\diff \Edist{x_i}{x_j}}
           =  \pi \weight{i} \sigma_{ij}
              \left( 1 - \frac{r_i^2-r_j^2}{\Edist{x_i}{x_j}^2} \right) , \\
  p_{ijk} &=  4 \pi \weight{i} r_i
              \frac{r_{ijk} \nu_{ijk}}{2 \pi r_i \Edist{x_i}{x_j}}
           =  2 \weight{i} r_{ijk} \nu_{ijk} \frac{1}{\Edist{x_i}{x_j}} ,
\end{align}
in which the first sum is over all directed boundary edges of the alpha shape,
and the second sum is over all triangles incident to these edges.
Observe that for fixed $i$, we get possibly non-zero contributions to all $\ppp_j$.
Symmetrically, we get $\ppp_i$ by accumulating contributions from all $j$.
Using $\uuu_{ji} = - \uuu_{ij}$ and $\uuu_{jik} = \uuu_{ijk}$, we get
\begin{align}
  \ppp_i  &=  \sum_j (p_{ij} + p_{ji}) \uuu_{ij}
         + \sum_{j,k} (p_{ijk} - p_{jik}) \uuu_{ijk} ,
  \label{eqn:pi}
\end{align}
in which the first sum is over all boundary edges of the alpha shape
incident to $x_i$,
and the second sum is over all triangles incident to these edges.
Not surprisingly, the result is similar to the weighted area gradient
given in Proposition \ref{prop:AreaDerivative}.
Specifically, we get $a_{ij} = p_{ij} r_i$ and $a_{ijk} = p_{ijk} r_i$.

\ourparagraph{Second term.}
We use Lemma \ref{lem:dRijPhiij} to rewrite \eqref{eqn:qprime} as
\begin{align}
      \!q'      &=  \sum_{i,j} \left( q_{ij} + \bar{q}_{ij} \right)
                               \scalprod{\uuu_{ij}}{\ttt_i-\ttt_j} ,         \\
      \!q_{ij}  &=  - \frac{\pi (\weight{i}+\weight{j})}{2}
                                    \phi_{ij} \sigma_{ij}
                      \frac{\diff r_{ij}}{\diff \Edist{x_i}{x_j}}            \\
                &=  \frac{- \pi (\weight{i}+\weight{j}) \phi_{ij} \sigma_{ij}
                      \left( (r_i^2 - r_j^2)^2 - \Edist{x_i}{x_j}^4 \right)}
                         {4 \Fdist{x_i}{x_j}^2 
                      \sqrt{ 2 \left( r_i^2 r_j^2 + (r_i^2+r_j^2)
                                      \Fdist{x_i}{x_j}^2 \right) 
                       \!-\! \left( r_i^4 + r_j^4 + \Fdist{x_i}{x_j}^4 \right)}} , \\
\!\bar{q}_{ij}  &=  - \frac{\pi (\weight{i}\!+\!\weight{j})}{2} r_{ij} \sigma_{ij}
                    \frac{\diff \phi_{ij}}{\diff \Fdist{x_i}{x_j}} 
                 =  \frac{- \pi (\weight{i}\!+\!\weight{j}) r_{ij} \sigma_{ij} \Edist{x_i}{x_j}}
                     {\!\!\sqrt{ 2 (r_i^2\!+\!r_j^2) \Fdist{x_i}{x_j}^2
                                 \!-\! (r_i^2\!-\!r_j^2)^2 - \Fdist{x_i}{x_j}^4 }} ,
\end{align}
in which the sum is over all directed boundary edges of the alpha shape.
Redistributing the terms, we get
\begin{align}
  \qqq_i  &=  \sum_j (q_{ij} + q_{ji} + \bar{q}_{ij} + \bar{q}_{ji}) \uuu_{ij} ,
  \label{eqn:qi}
\end{align}
in which the sum is over all boundary edges of the alpha shape
incident to $x_i$.

\ourparagraph{Third term.}
The redistribution of the pieces on the right-hand side of
\eqref{eqn:dsigmaij} to get the gradient is complicated by the change
of the motion.
We therefore begin by rewriting $\TTT_{ijk}$.
To this end, we recall that the cross product is distributive:
$a \times b + a \times c = a \times (b+c)$,
and that the Lagrange formula turns a triple cross product
into two scalar products:
$(a \times b) \times c = b \scalprod{a}{c} - a \scalprod{b}{c}$.
Recalling the definition of $D$ from \eqref{eqn:D},
we can now rewrite the new motion vector:
\begin{align}
  \TTT_{ijk}  &=  \ttt_k - \ttt_i - \ooomega_j \times (x_k - x_i)
                - D [\ttt_j - \ttt_i - \ooomega_j \times (x_j - x_i) ]  \\
              &=  \ttt_k - D \ttt_j + (D-1) \ttt_i
                - \ooomega_j \times [ x_k - D x_j + (D-1) x_i ]         \\
              &=  \ttt_k - D \ttt_j + (D-1) \ttt_i
                - \scalprod{\ttt_j-\ttt_i}{\ddd_{ijk}} \uuu_{ij}
                + \scalprod{\uuu_{ij}}{\ddd_{ijk}} (\ttt_j-\ttt_i) ,
\end{align}
in which the third line is obtained by applying the Lagrange formula
and writing $\ddd_{ijk} = (x_k - D x_j + (D-1) x_i) / \Edist{x_i}{x_j}$.
We need the scalar product of $\TTT_{ijk}$ with $x_k - P$:
\begin{align}
  \scalprod{\TTT_{ijk}}{x_k-P}
    &=  \scalprod{x_k-P}{\ttt_k - D \ttt_j + (D-1) \ttt_i}                 \\
    &~~~~ - \scalprod{x_k-P}{\uuu_{ij}} \scalprod{\ddd_{ijk}}{\ttt_j - \ttt_i} 
      + \scalprod{\uuu_{ij}}{\ddd_{ijk}} \scalprod{x_k-P}{\ttt_j - \ttt_i} \\
    &=  \scalprod{\aaa_{ijk}}{\ttt_k}
      + \scalprod{\bbb_{ijk}}{\ttt_j}
      + \scalprod{\ccc_{ijk}}{\ttt_i} , 
\end{align}
in which
\begin{align}
  \aaa_{ijk}  &=  x_k - P ,                                                \\
  \bbb_{ijk}  &=  [ - D + \scalprod{\uuu_{ij}}{\ddd_{ijk}} ] (x_k-P)
                - \scalprod{x_k-P}{\uuu_{ij}} \ddd_{ijk} ,                 \\
  \ccc_{ijk}  &=  [D-1 - \scalprod{\uuu_{ij}}{\ddd_{ijk}} ] (x_k-P)
                + \scalprod{x_k-P}{\uuu_{ij}} \ddd_{ijk} .
\end{align}
We are now ready to rewrite \eqref{eqn:sprime}
using Lemma \ref{lem:dSigmaij} as
\begin{align}
  s'    &=  \sum_{i, j, P}
              \left[ s_{ijk} \scalprod{\TTT_{ijk}}{x_k-P}
               + \bar{s}_{ijk} \scalprod{\uuu_{ij}}{\ttt_j-\ttt_i} \right] , \\
        &=  \sum_{i, j, P}
              \left[ s_{ijk} \scalprod{\aaa_{ijk}}{\ttt_k}
                   \!+\! s_{ijk} \scalprod{\bbb_{ijk}}{\ttt_j}
                   \!+\! s_{ijk} \scalprod{\ccc_{ijk}}{\ttt_i}
                   \!+\! \bar{s}_{ijk} \scalprod{\uuu_{ij}}{\ttt_j}
                   \!-\! \bar{s}_{ijk} \scalprod{\uuu_{ij}}{\ttt_i} \right] ,  \\
  s_{ijk}        &=  - \frac{\weight{i}+\weight{j}}{4} \phi_{ij}
                       \frac{1}{\scalprod{x_k-P}{\uuu_{ij}^P}} ,                         \\
  \bar{s}_{ijk}  &=    \frac{\weight{i}+\weight{j}}{4} \phi_{ij}
                       \frac{\diff \Aalpha{P}}{\diff r_{ij}} 
                       \frac{\diff r_{ij}}{\diff \Edist{x_i}{x_j}} ,
\end{align}
in which the sum is over the three ordered versions of all oriented
boundary triangles of the alpha shape.
Letting $i, j, k$ be the vertices of such a triangle,
the ordered versions are given by the index triplets $ijk, kij, jki$,
and we write $P$ for the corresponding corner of the space-filling diagram,
noting that $P \in S_{ijk}$.
Observe that an unordered triangle that belongs to $0, 1, 2$ tetrahedra in the
alpha complex occurs $2, 1, 0$ times in this sum.
Redistributing the terms, we get
\begin{align}
  \sss_i  &=  \sum_{j, k} \left[
              s_{ijk} \aaa_{ijk} + s_{kij} \bbb_{kij} + s_{jki} \ccc_{jki}
                  + \bar{s}_{jik} \uuu_{ji} - \bar{s}_{ijk} \uuu_{ij} \right] ,
  \label{eqn:si}
\end{align}
in which the sum is over all oriented boundary triangles of the alpha shape
that share $x_i$.
As before, $P \in S_{ijk}$ is the corresponding corner of the space-filling diagram.

\ourparagraph{Summary.}
We finally get the gradient of the weighted mean curvature function by adding
the gradients of the three component functions:
\begin{theorem}[Gradient of Weighted Mean Curvature]
  \label{thm:WMCG}
  The derivative of the weighted mean curvature
  of the space-filling diagram at state $\xxx$ with momentum $\ttt$ is
  $\Diff \mean_\xxx (\ttt) = \scalprod{\mmm}{\ttt}$,
  in which $\mmm_i  =  \ppp_i + \qqq_i + \sss_i$ as given in
  \eqref{eqn:pi}, \eqref{eqn:qi}, and \eqref{eqn:si}, for all $0 \leq i < n$.
\end{theorem}

\section{Violations of Continuity}
\label{sec:5}

To embed our formulas in the inner loop of a molecular dynamics application,
the implementation must be efficient and robust.
We address the latter requirement by identifying the
subset of $\Rspace^{3n}$
where the gradient of the weighted mean curvature function is not continuous.
This subset is contained in the subset of non-generic states,
which we describe first.

\ourparagraph{General position.}
We distinguish two types of degenerate states:
where the Delaunay mosaic is ambiguous,
and where the Delaunay mosaic is unambiguous but the alpha complex is ambiguous.
Recall that $X$ is a collection of $n$ closed balls in $\Rspace^3$,
and that these balls can move individually but their radii are fixed.
We say $X$ is in \emph{general position} and, equivalently,
that its state is \emph{generic} if the following three conditions hold:
\medskip \begin{enumerate}\denselist
  \item[I.]  the common intersection of $p+1$ Voronoi domains is either empty
    or a convex polyhedron of dimension $3-p$;
  \item[II.] the common intersection of $p+1$ spheres bounding balls in $X$
    is either empty or a sphere of dimension $2-p$.
\end{enumerate} \medskip
Condition I implies that any five Voronoi domains have an empty common intersection,
and Condition II implies that any four spheres have an empty common intersection.
Each violation of Condition I corresponds to
a $(3n-1)$-dimensional submanifold of $\Rspace^{3n}$
and so does every violation of Condition II, except when two radii are the same,
in which case we get a submanifold of dimension $3n-3$.
Let $\Mspace{I}$ and $\Mspace{II}$ be the union of submanifolds
that correspond to violations of Conditions I and II, respectively.
Since there are only finitely many such submanifolds,
$\Mspace{I}$ and $\Mspace{II}$ have dimension $3n-1$ each.

\ourparagraph{Condition I and flips.}
If Condition I is satisfied, then the Delaunay mosaic is simplicial.
We get a violation when the state trajectory intersects $\Mspace{I}$.
We limit ourselves to discussing what we call the \emph{typical case},
in which there is only one violation of general position.
Indeed, multiple violations correspond to intersections of the
$(3n-1)$-dimensional submanifolds.
In the typical case, the intersection is an isolated point where
the trajectory passes locally from one side of $\Mspace{I}$ to the other.
The corresponding change in the Delaunay mosaic is a flip,
of which there are four types.
For integers $1 \leq b \leq 4$ and $a = 5 - b$,
the \emph{$b$-to-$a$ flip} replaces $b$ tetrahedra by $a$ tetrahedra;
see \cite{Ede01} for details on this operation.
The Delaunay mosaic is a simplicial complex geometrically realized in $\Rspace^3$,
both before and after the replacement.
This implies that the union of the $b$ tetrahedra before the flip
be the same as the union of the $a$ tetrahedra after the flip.

To be more formal, let $B$ be the complex consisting of the $b$ tetrahedra
and their faces, and let $A$ be the complex consisting of the $a$ tetrahedra
and their faces.
Then $C = B \cap A$ is the common boundary of both complexes,
and the flip substitutes $A \setminus C$ for $B \setminus C$ in
the Delaunay mosaic.
By construction, either all simplices of $B \setminus C$
and of $A \setminus C$ belong to the
alpha complex before and after the flip, or none does.
In the latter case, none of the formulas in Theorem \ref{thm:WMCG} are affected
by the flip.
In the former case, some terms in the formulas get replaced by other terms.
At the moment of the flip, the replaced and the replacing terms
evaluate to the same numerical value,
which implies that the weighted mean curvature and its derivative are
unchanged and therefore continuous.
We can see this by noting that the flip does not alter
the combinatorial structure of the space-filling diagram boundary
and therefore affects neither the weighted mean curvature nor its derivative.

\ourparagraph{Condition II and critical simplices.}
Condition II addresses situations in which the alpha complex changes while
the Delaunay mosaic stays the same:
we either add one or more simplices in the mosaic to the complex or
we remove them from the complex.
Here we consider the case in which only one simplex is added or removed.
With reference to the discrete Morse theory of Delaunay mosaics
outlined in \cite{BaEd17}, we call this a \emph{critical simplex}.
We distinguish three cases.
\medskip \begin{description}\denselist
  \item[{\sc Case} $\Case{C}{1}$:]  the added or removed simplex is an edge
    of the Delaunay mosaic.
    At the moment of the change, the spheres whose centers are the endpoints of
    the edge touch in a single point.
    Letting $i$ and $j$ be the indices of the spheres,
    $\ee = r_i + r_j - \Edist{x_i}{x_j}$ is negative when the spheres are disjoint
    and positive when they intersect in a circle.
    For $\ee < 0$, the edge does not belong to the alpha complex and therefore
    contributes neither to the weighted mean curvature nor to its gradient.
    For $\ee > 0$, the change to the weighted mean curvature caused by the edge
    depends on the change in area of the space-filling diagram,
    the change in length of its arcs, and the dihedral angles at the arcs.
    Focussing on the rough order of the changes,
    it is not difficult to see that
    $\Delta \area = \ee$, $\Delta \length = \sqrt{\ee}$, $\Angle = 1$,
    and therefore $\Delta \mean = \sqrt{\ee}$ and $\norm{\nabla \mean} = \infty$
    at $\ee = 0$.
\end{description} \medskip
The other two cases are similar so we will be brief.
In each case, we use $\ee$ to parameterize the local motion,
with $\ee = 0$ at the moment of change.
We feel free to make further non-essential simplifying assumptions,
such as the radii being distinct.
 
\medskip \begin{description}\denselist
  \item[{\sc Case} $\Case{C}{2}$:]  the added or removed simplex is a triangle
    of the Delaunay mosaic.
    At the moment of change, the spheres whose centers are the vertices of the
    triangle meet in a single point.
    By easy analysis, we get
    $\Delta \area = \ee \sqrt{\ee}$, $\Delta \length = \sqrt{\ee}$,
    $\Angle = 1$,
    and therefore $\Delta \mean = \sqrt{\ee}$ and $\norm{\nabla \mean} = \infty$
    at $\ee = 0$.
    
  \item[{\sc Case} $\Case{C}{3}$:]  the added or removed simplex is a tetrahedron
    of the Delaunay mosaic.
    At the moment of change, the spheres whose centers are the vertices of the
    tetrahedron meet in a single point.
    By easy analysis, we get
    $\Delta \area = \ee^2$, $\Delta \length = \ee$, $\Angle = 1$,
    and therefore $\Delta \mean = \ee$ and $\Delta \norm{\nabla \mean} = 1$
    at $\ee = 0$.
\end{description} \medskip
In conclusion, the weighted mean curvature is continuous but its gradient
occasionally changes discontinuously;
see Table \ref{tbl:changes} for a convenient summary
of the three cases above as well as the six cases to be discussed shortly.
\begin{table}[hbt]
  \centering
  \begin{tabular}{c || ccc | cccccc}
                                  & $\Case{C}{1}$  & $\Case{C}{2}$  & $\Case{C}{3}$
                                  & $\Case{N}{01}$ & $\Case{N}{02}$ & $\Case{N}{03}$ & $\Case{N}{12}$ & $\Case{N}{13}$ & $\Case{N}{23}$  \\
    \hline \hline
    $\Delta \mean$                & $\sqrt{\ee}$   & $\sqrt{\ee}$   & $\ee$       
                                  & $\ee$          & $\sqrt{\ee}$   & $\ee$          & $\sqrt{\ee}$   & $\ee$          & $\ee$           \\
    $\Delta \norm{\nabla \mean}$  & $\infty$       & $\infty$       & $1$           
                                  & $1$            & $\infty$       & $1$            & $\infty$       & $1$            & $1$  
  \end{tabular}
  \caption{The order of change of the weighted mean curvature and of its
    gradient at states that violate Condition II.}
  \vspace{-0.2in}
  \label{tbl:changes}
\end{table}

\Skip{
  \item  An edge is added or removed when two spheres kiss:
    points $x_i$ and $x_j$ with distance $\Edist{x_i}{x_j} = r_i + r_j$.
    To simplify the analysis, we assume $\weight{i} = \weight{j} = 1$
    and $r_i = r_j = r$.
    When the spheres move closer, to distance $\Edist{x_i}{x_j} - 2 \ee$,
    the surface area shrinks by $4 \pi r \ee$.
    Accordingly, the surface part of the mean curvature shrinks by $4 \pi \ee$.
    At the same time, the mean curvature grows by the contribution of
    the intersection circle, which is $2 \pi r_{ij} \phi_{ij}$.
    For small $\ee > 0$, we can approximate $r_{ij}$ by $\sqrt{2 r \ee}$
    and $\phi_{ij}$ be $\pi$,
    which implies that the growth is approximated by a constant times $\sqrt{\ee}$.
    Accordingly, the derivative jumps from $0$ before the sphere touch
    to $- \infty$ at the moment they meet.
}

\ourparagraph{Condition II and non-singular intervals.}
Here we consider the case in which at least two simplices are added or
removed at the same moment.
Generically, these simplices form an interval of size $2$, $4$, or $8$;
see \cite{BaEd17} for details.
We encounter six cases.
\medskip \begin{description}\denselist
  \item[{\sc Case} $\Case{N}{01}$:]
    A vertex together with an incident edge are added or removed.
    At the moment of change, the sphere centered at the vertex
    breaks through the surface of the sphere whose center is the other
    endpoint of the edge.
    By easy analysis, we get
    $\Delta \area = \ee$, $\Delta \length = \sqrt{\ee}$,
    and because $\phi_{ij}$ is proportional to $r_{ij}$
    also  $\Angle = \sqrt{\ee}$.
    Hence $\Delta \mean = \ee$ and $\Delta \norm{\nabla \mean} = 1$
    at $\ee = 0$.

  \item[{\sc Case} $\Case{N}{02}$:]
    A vertex together with an incident triangle and its two edges that share
    the vertex are added or removed.
    At the moment of change, the sphere centered at the vertex breaks through
    the circle at which the two spheres centered at the other vertices
    of the triangle intersect.
    By easy analysis, we get
    $\Delta \area = \ee \sqrt{\ee}$, $\Delta \length = \sqrt{\ee}$,
    $\Angle = 1$,
    and therefore $\Delta \mean = \sqrt{\ee}$ and $\norm{\nabla \mean} = \infty$
    at $\ee = 0$.

  \item[{\sc Case} $\Case{N}{03}$:]
    A vertex together with an incident tetrahedron and its three edges
    and three triangles that share the vertex are added or removed.
    At the moment of change, the sphere centered at the vertex breaks through
    the pair of points in which the three spheres centered at the other vertices
    of the tetrahedron intersect.
    We get $\Delta \area = \ee^2$, $\Delta \length = \ee$,
    $\Angle = 1$,
    and therefore $\Delta \mean = \ee$ and $\Delta \norm{\nabla \mean} = 1$
    at $\ee = 0$.
 
  \item[{\sc Case} $\Case{N}{12}$:]
    An edge together with an incident triangle are added or removed.
    At the moment of change, the circle in which the two spheres centered
    at the endpoints of the edge intersect breaks through the surface
    of the sphere centered at the third vertex of the triangle.
    By easy analysis, we get
    $\Delta \area = \ee \sqrt{\ee}$, $\Delta \length = \sqrt{\ee}$,
    $\Angle = 1$,
    and therefore $\Delta \mean = \sqrt{\ee}$
    and $\norm{\nabla \mean} = \infty$ at $\ee = 0$.

  \item[{\sc Case} $\Case{N}{13}$:]
    An edge together with an incident tetrahedron and its two triangle
    that share the edge are added or removed.
    At the moment of change, the circle in which the two spheres centered
    at the endpoints of the edge intersect
    breaks through the circle in which the two spheres centered at the
    other vertices of the tetrahedron intersect.
    We get $\Delta \area = \ee^2$, $\Delta \length = \ee$,
    $\Angle = 1$,
    and therefore $\Delta \mean = \ee$ and $\Delta \norm{\nabla \mean} = 1$
    at $\ee = 0$.

  \item[{\sc Case} $\Case{N}{23}$:]
    A triangle together with an incident tetrahedron are added or removed.
    At the moment of change, the pair of points in which the three spheres centered
    at the vertices of the triangle intersect
    break through the surface of the sphere centered at the fourth vertex
    of the tetrahedron.
    We get $\Delta \area = \ee^2$, $\Delta \length = \ee$,
    $\Angle = 1$,
    and therefore $\Delta \mean = \ee$ and $\Delta \norm{\nabla \mean} = 1$
    at $\ee = 0$.
\end{description}\medskip
See again Table \ref{tbl:changes} for a convenient summary of the results
in all cases.
We note that the situation in the unweighted case has better continuity
properties along $\Mspace{II} \subseteq \Rspace^{3n}$ in some cases,
but the analysis is more involved.
Subtracting $\Mspace{I} \cup \Mspace{II}$ from $\Rspace^{3n}$,
we are left with finitely many open cells such that the formula of
the weighted mean curvature and of its derivative are both invariant
over each cell.
All terms in these formulas are continuous over the cell,
which implies that both $\mean \colon \Rspace^{3n} \to \Rspace$
and $\nabla \mean \colon \Rspace^{3n} \to \Rspace^{3n}$
are continuous over the open cell.
As argued above, $\mean$ and $\nabla \mean$ are also continuous
at states $\xxx \in \Mspace{I} \setminus \Mspace{II}$,
hence they are continuous at all
$\xxx \in \Rspace^{3n} \setminus \Mspace{II}$.
We summarize the findings of this section.
\begin{theorem}[Continuity of Gradient]
  \label{thm:CofG}
  The gradient of the weighted mean curvature of a space-filling diagram
  of $n$ closed balls in $\Rspace^3$ is continuous provided
  the state $\xxx \in \Rspace^{3n}$ of the diagram does not belong
  to $\Mspace{II}$, which is a $(3n-1)$-dimensional
  subset of $\Rspace^{3n}$.
\end{theorem}

\section{Discussion}
\label{sec:6}

The main contribution of this paper is the analysis of the derivative
of the weighted mean curvature of the space-filling diagram
of a set of moving balls.
Specifically, we give an explicit description of the gradient of
the weighted mean curvature function, which for $n$ spheres is
a map from $\Rspace^{3n}$ to $\Rspace$.
In addition, we characterize the subset of $\Rspace^{3n}$ at which
the derivative violates continuity.
In total, this is sufficient information for an efficient and robust
implementation of the weighted mean curvature derivative,
one that can be added to the inner loop of a molecular dynamics simulation
of a physical system.
There are several questions about this application that go beyond
the scope of this paper:
\medskip \begin{itemize}\denselist
  \item  Is there a connection between the coefficients with which the
    morphological approach combines the intrinsic volumes \cite{HRC13,RHK06}
    and the states at which their gradients are not continuous?
  \item  Splitting the mean curvature concentrated along an arc in equal parts
    is suggested by the Apollonius diagram of the spheres, but splitting it
    according to the Voronoi diagram is also feasible.
    Are there physical reasons to prefer one split over the other?
\end{itemize} \medskip
We remark that the formulas in this paper can be easily adapted to splitting
the mean curvature according to the Voronoi tessellation of the spheres.

\subsection*{Acknowledgment}
\footnotesize{
The authors of this paper thank Roland Roth for suggesting the analysis of
the weighted curvature derivatives for the purpose of improving molecular
dynamics simulations and for his continued encouragement.
They also thank Patrice Koehl for the implementation of the formulas
and for his encouragement and advise along the road.
Finally, they thank two anonymous reviewers for their constructive criticism.
}


\appendix \clearpage
\section{Notation}
\label{app:N}

\begin{table}[hbt]
  \centering \begin{tabular}{ll}
    $S_i = \boundary{B_i}$
      &  sphere bounding ball                                           \\
    $S_{ij} = \boundary{B_{ij}}$
      &  circle bounding disk                                           \\
    $S_{ijk} = \boundary{B_{ijk}}$
      &  pair of points bounding line segment                           \\
    $x_i, x_{ij}, x_{ijk}$
      &  centers of $S_i$, $S_{ij}$, $S_{ijk}$                          \\
    $r_i, r_{ij}, r_{ijk}$
      &  radii of $S_i$, $S_{ij}$, $S_{ijk}$                            \\
    $\uuu_{ij} = \frac{x_i - x_j}{\Edist{x_i}{x_j}}$
      &  unit vector between centers                                    \\
    $\uuu_{ijk} = \tfrac{\uuu_{ik} - \scalprod{\uuu_{ik}}{\uuu_{ij}} \uuu_{ij}}
                  {\|{\uuu_{ik} - \scalprod{\uuu_{ik}}{\uuu_{ij}} \uuu_{ij}}\|}$
      &  unit normal to $\uuu_{ij}$ with positive component in direction $\uuu_{ik}$  \\
    $\nu_i = \frac{\Volume(B_i \cap \Vdom{i})}{\Volume(B_i)}$
      &  volume fraction of ball                                        \\
    $\nu_{ij} = \frac{\Area(B_{ij} \cap \Vdom{ij})}{\Area(B_{ij})}$
      &  area fraction of disk                                          \\
    $\nu_{ijk} = \frac{\Length(B_{ijk} \cap \Vdom{ijk})}{\Length(B_{ijk})}$
      &  length fraction of line segment                                \\
    $\nu_{ijk\ell} = \frac{\Card{B_{ijk\ell} \cap \Vdom{ijk\ell}}}
                          {\Card{B_{ijk\ell}}}$
      &  $0$ or $1$                                                     \\
    $\sigma_i = \frac{\Area(S_i \cap \Vdom{i})}{\Area(B_i)}$
      &  area fraction of sphere                                        \\
    $\sigma_{ij} = \frac{\Length(S_{ij} \cap \Vdom{ij})}{\Length(S_{ij})}$
      &  length fraction of circle                                      \\
    $\sigma_{ijk} = \frac{\Card{S_{ijk} \cap \Vdom{ijk}}}{\Card{S_{ijk}}}$
      &  $0$, $\tfrac{1}{2}$, or $1$                                    \\
                                                                        \\
    $W = \Welec + \Wnp$
      &  effective solvation potential                            \\
    $K, K_r = K \oplus r \Bspace^3$
      &  convex body, parallel body                               \\
    $H \in \Grass{i}{3}, \Euler{K \cap H}$
      &  $i$-plane, Euler characteristic of intersection          \\
    $X = \{ B_i \}, \bigcup X$
      &  set of balls, space-filling diagram                      \\
    $\Sspace^2, \Bspace^3, \Rspace^3; f \colon \Sspace^2 \to \Rspace$
      &  unit sphere, unit ball, Euclidean space; height function \\
    $\xxx, \ttt; \vvv, \aaa$
      &  state, momentum; gradients of volume, area               \\
    $F \colon \Rspace^{3n} \to \Rspace$
      &  intrinsic volume function                                \\
    $v_{ij}, \bar{v}_{ij}, a_{ij}, a_{ijk}$
      &  constants                                                \\
    $x, a, b, c, \gamma$
      &  abstract variables                                       \\
    $\xi_i, \xi_j$
      &  signed distances                                         \\
    $\ttt_k, \TTT_{ijk}$
      &  velocity vectors                                         \\
    $\ooomega_j; \tttStr_{j}, \tttRot_{j}$
      &  angular momentum; components of motion                   \\
    $A; P; v; \uuu_{ij}^P; \Aalpha{P}$
      &  area; point; speed; vector; angle                        \\
    $\VVV_{ij}, \VVV_{ijk}, v_k, h_k$
      &  auxiliary variables                                      \\
    $\mean' = p' + q' + s'$
      &  derivatives                                              \\
    $\ppp, \qqq, \sss; \ppp_i, \qqq_i, \sss_i$
      &  gradients; as they apply to $x_i$                        \\
    $\aaa_{ijk}, \bbb_{ijk}, \ccc_{ijk}, \ddd_{ijk}$
      &  auxiliary vectors                                         \\
    $p_{ij}, p_{ijk}, q_{ij}, \bar{q}_{ij}, s_{ijk}, \bar{s}_{ijk}$
      &  constants                                                \\
    $\angle x_ix_j, \angle x_ix_jx_k; \ee$
      &  angle parameters; motion parameter                       \\
    $\Mspace{I}, \Mspace{II}$
      &  subsets of $\Rspace^{3n}$ 
  \end{tabular}
  \caption{Notation for concepts, sets, functions, vectors, variables.}
  \label{tbl:Notation}
\end{table}

\newpage
\section{Definitions and Results}
\label{app:DR}

\begin{itemize}\denselist
  \item Section \ref{sec:1}: Introduction.
  \item Section \ref{sec:2}: Background.
    \begin{itemize}\denselist
      \item Formula \ref{form:Archimedes} (Archimedes' Theorem).
      \item Formula \ref{form:Heron} (Heron's Theorem).
      \item Proposition \ref{prop:IntrinsicVolumes} (Intrinsic Volumes).
      \item Proposition \ref{prop:VolumeDerivative} (Gradient of Weighted Volume).
      \item Proposition \ref{prop:AreaDerivative} (Gradient of Weighted Area).
    \end{itemize}
  \item Section \ref{sec:3}: Derivatives.
    \begin{itemize}\denselist
      \item Lemma \ref{lem:dSigmai} (Derivative of $\sigma_i$).
      \item Lemma \ref{lem:dRijPhiij} (Derivatives of $r_{ij}$ and of $\phi_{ij}$).
      \item Lemma \ref{lem:ReMotion} (Change of Motion).
      \item Lemma \ref{lem:dSigmaij} (Derivative of $\sigma_{ij}$).
    \end{itemize}
  \item Section \ref{sec:4}: Gradients.
    \begin{itemize}\denselist
      \item Theorem \ref{thm:WMCG} (Gradient of Weighted Mean Curvature).
    \end{itemize}
  \item Section \ref{sec:5}: Violations of Continuity.
    \begin{itemize}\denselist
      \item Theorem \ref{thm:CofG} (Continuity of Gradient).
    \end{itemize}
  \item Section \ref{sec:6}: Discussion.
\end{itemize}

\end{document}